\documentclass[journal,10pt]{IEEEtran}

\usepackage{amsfonts}
\usepackage{amsmath}
\usepackage{graphicx}
\usepackage{color}
\usepackage{multicol}
\usepackage{amssymb}
\usepackage{subfigure}
\usepackage{bm}
\usepackage{latexsym}
\usepackage{stfloats}
\usepackage{float}
\usepackage{exscale}
\usepackage{relsize}
\usepackage{cite}
\usepackage{amsthm,amsmath}
\usepackage[ruled,linesnumbered]{algorithm2e}
\usepackage{mathtools}
\usepackage{xcolor}

\begin{document}

\title{Analysis and  Optimization for RIS-Aided
Multi-Pair Communications Relying on Statistical CSI}

\author{Zhangjie~Peng,
        Tianshu~Li,
        Cunhua Pan,~\IEEEmembership{Member,~IEEE,}
        Hong~Ren~\IEEEmembership{Member,~IEEE,}
         \\
        Wei~Xu,~\IEEEmembership{Senior~Member,~IEEE,}
        and~Marco Di Renzo~\IEEEmembership{Fellow,~IEEE}
        % <-this % stops a space
\vspace{-0.0cm}

\thanks{
\emph{(Corresponding authors: Cunhua Pan.)}}
\thanks{Z. Peng is with the College of Information, Mechanical and Electrical Engineering,
 Shanghai Normal University, Shanghai 200234, China,
 and also with the Institute of Artificial Intelligence
on Education, Shanghai Normal University, Shanghai 200234, China (e-mail: pengzhangjie@shnu.edu.cn).}
\thanks{T. Li is with the College of Information, Mechanical and Electrical Engineering,
Shanghai Normal University, Shanghai 200234, China $( \text{e-mail: 1000479056@smail.shnu.edu.cn} )$.}
\thanks{C. Pan and H. Ren are with the School of Electronic Engineering and Computer Science at Queen
Mary University of London, London E1 4NS, U.K. $( \text{e-mail: {c.pan,h.ren}@qmul.ac.uk} )$.}
\thanks{W. Xu is with the National Mobile Communications Research Lab, Southeast University, Nanjing 210096, China, and also with Henan Joint International Research Laboratory of Intelligent Networking and Data Analysis, Zhengzhou University, Zhengzhou, 450001 China (wxu@seu.edu.cn).}
\thanks{M. Di Renzo is with Universit\'e Paris-Saclay, CNRS, CentraleSup\'elec, Laboratoire des Signaux et Syst\`emes, 3 Rue Joliot-Curie, 91192 Gif-sur-Yvette, France. (marco.di-renzo@universite-paris-saclay.fr)}
\vspace{-0.2cm}
}

\maketitle

\newtheorem{lemma}{Lemma}
\newtheorem{theorem}{Theorem}
\newtheorem{remark}{Remark}

\begin{abstract}
In this paper, we investigate a reconfigurable intelligent surface (RIS) aided multi-pair communication system, in which multi-pair users exchange information via an RIS. We derive an approximate expression for the achievable rate by assuming that statistical channel state information (CSI) is available. A genetic algorithm (GA) to solve the rate maximization problem is proposed as well.
In particular, we consider implementations of RISs with continuous phase shifts (CPSs) and discrete phase shifts (DPSs).
Simulation results verify the obtained results and show that the proposed GA method has almost the same performance as the globally optimal solution.
In addition, numerical results show that three quantization bits can achieve a large portion of the achievable rate for the CPSs setup.
\end{abstract}

\begin{IEEEkeywords}
Reconfigurable intelligent surface (RIS),
intelligent reflecting surface (IRS),
statistical channel state information (CSI),
multi-pair communication,
genetic algorithm (GA).
\end{IEEEkeywords}

\IEEEpeerreviewmaketitle

\section{Introduction}

Reconfigurable intelligent surfaces (RISs) are an emerging transmission technology that is capable of configuring the wireless channel into desirable forms by appropriately optimizing the response of their individual scattering elements \cite{9140329}.
Due to their appealing features of low cost and low power consumption, RISs have attracted extensive research attention \cite{8741198, 8811733, 2019arXiv191112296H, 2019arXiv190308925D}.

Some initial efforts have been devoted to the study of  various RIS-aided communication systems and applications, such as physical layer security in \cite{Shen2019Secrecy}, \cite{2019arXiv191201497Y}, multicell networks in \cite{9090356}, full duplex systems in \cite{200605147}, mobile edge computing in \cite{Tong2020Latency}, and wireless power transfer in \cite{2019arXiv190804863P}.
%More specifically, for the secure communication systems, the authors of \cite{Shen2019Secrecy} considered
%physical layer security design for the RIS-assisted multi-antenna system.
%Additionally, the secrecy rate is maximized by jointly optimizing the active beamforming and the passive phase shift in \cite{2019arXiv191201497Y} .
%For the multicell network, the performance of the cell-edge user can be enhanced by employing an RIS at the cell boundary to mitigate the adjacent-cell interference \cite{9090356}.
%A FD two-way communication was investigated in \cite{200605147}, where the authors demonstrated that deploying an RIS near the base station is able to cover the dead zone in the cell.
%In \cite{Tong2020Latency}, the RIS was utilized to help in reducing the latency of MEC networks.
%In addition, the authors of \cite{2019arXiv190804863P} showed that the RIS is beneficial in enlarging the operating region of the energy harvesting sensors.
However, a paucity of contribution studied  RIS-aided multi-pair communication systems, which is a typical communication scenario due to the rapid increase of the number of machine-type devices in future networks.

In multi-pair communication systems, the direct communication links may be blocked in both indoor and outdoor scenarios. In particular, the direct signals may be blocked by trees and large buildings in outdoor scenarios. In indoor scenarios, the signals may be blocked by thick walls, especially in high-frequency millimeter-wave communication systems.
To counteract these issues, we consider the deployment of an RIS in order to assist the transmission between pairs of devices.
Compared with existing research works on relay-assisted multi-pair communications in \cite{9106840,7947159}, RISs have some appealing advantages. In particular, an RIS has low power consumption and does not need power amplifiers and signal processing units after configuration \cite{9140329, 2019arXiv190808747D}.

\begin{figure}[t]
\centering
\includegraphics[scale=0.28]{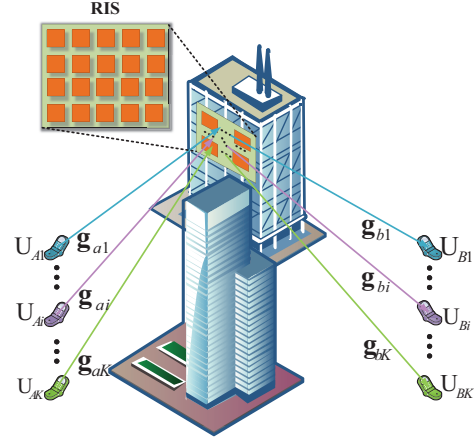}
\vspace{-0.4cm}
\caption{System model for RIS-aided multi-pair communications.}
\label{multipair}
\vspace{-0.4cm}
\end{figure}

Against this background, we study the transmission design for an RIS-aided multi-pair communication system. Unlike most of the existing papers \cite{Shen2019Secrecy,2019arXiv191201497Y,9090356,200605147,Tong2020Latency,2019arXiv190804863P} where instantaneous channel  state information (CSI) is assumed to be known, we consider the availability of statistical CSI \cite{9005237} that is easier to obtain since it varies more slowly.
Specifically, our contributions are as follows: 1) We derive the achievable rate of the considered system model; 2)
we propose a genetic algorithm (GA) method to solve the phase shifts optimization problem, by considering the setups of continuous phase shifts (CPSs) and discrete phase shifts (DPSs); 3) we provide extensive simulation results to demonstrate the correctness of our derived results, and to show that three bits are enough to discretize the phase shifts, which provides useful engineering insights for designing RIS-aided systems.

The rest of the paper is organized as follows.
In Section
\uppercase\expandafter{\romannumeral2}, we introduce the RIS-aided multi-pair communication system model.
We derive the achievable rate in Section \uppercase\expandafter{\romannumeral3} and optimize the phase shifts in Section \uppercase\expandafter{\romannumeral4}.
Numerical results are provided to demonstrate the correctness of our analysis in Section \uppercase\expandafter{\romannumeral5}.
Conclusions are drawn in Section \uppercase\expandafter{\romannumeral6}.

\section{System Model}

We consider an RIS-aided multi-pair communication system, where $K$ pairs of users exchange information via an RIS, as shown in Fig. \ref{multipair}.
The RIS consists of $L$ reflective elements that are capable of customizing the channel by optimizing their phase shifts.
The phase shift matrix $\bm{\Theta}$ is denoted by $\bm{\Theta}=diag(e^{j\theta_1},\ldots,e^{j\theta_\ell},\ldots,e^{j\theta_L})$, where $\theta_\ell$ is the phase shift of the $\ell$th reflective element.
We denote the $i$th single-antenna transmitter as $\text U_{Ai}$ and the $i$th single-antenna receiver as $\text U_{Bi}$, for $i = 1, . . . , K$.

The channel between $\text U_{Ai}$ and the RIS and the channel between the RIS and $\text U_{Bi}$ can be written as
\vspace{-0.2cm}
\begin{equation}
\vspace{-0.1cm}
\textbf{g}_{ai} =\sqrt{\alpha_{ai}}\textbf{h}_{ai},
\end{equation}
\begin{equation}
\vspace{-0.2cm}
\textbf{g}_{bi} =\sqrt{\alpha_{bi}}\textbf{h}_{bi},
\end{equation}
where $\alpha_{ai}$ and $\alpha_{bi}$ denote the large-scale fading coefficients, and
$\textbf{g}_{i} \in {{\mathbb C}^{L \times 1}}$ and $\textbf{h}_{i} \in {{\mathbb C}^{L \times 1}}$ denote the fast fading vectors.
Rician fading is assumed for all channels,
thus
the vectors $\textbf{h}_{ai}$ and $\textbf{h}_{bi}$ can be expressed as
\vspace{-0.2cm}
\begin{equation}\label{gi}
\vspace{-0.1cm}
\textbf{h}_{ai} =\sqrt{\frac{\varepsilon_i}{\varepsilon_i+1}}\overline{\textbf{h}}_{ai}+\sqrt{\frac{1}{\varepsilon_i+1}} \tilde{\textbf{h}}_{ai},
\end{equation}
\begin{equation}\label{hi}
\vspace{-0.2cm}
\textbf{h}_{bi} =\sqrt{\frac{\beta_i}{\beta_i+1}}\overline{\textbf{h}}_{bi}+\sqrt{\frac{1}{\beta_i+1}} \tilde{\textbf{h}}_{bi},
\end{equation}
where $\varepsilon_i$ and $\beta_i$ denote the Rician factors,
$\tilde{\textbf{h}}_{ai} \in {{\mathbb C}^{L \times 1}}$ and $\tilde{\textbf{h}}_{bi} \in {{\mathbb C}^{L \times 1}}$ denote the non-line-of-sight channel vectors,
whose entries are independent and identically distributed standard Gaussian random variables, i.e., $\mathcal{CN}(0,1)$,
and
$\overline{\textbf{h}}_{ai} \in {{\mathbb C}^{L \times 1}}$ and $\overline{\textbf{h}}_{bi} \in {{\mathbb C}^{L \times 1}}$ denote the line-of-sight channel vectors. In particular, $\overline{\textbf{h}}_{ai}$ and $\overline{\textbf{h}}_{bi}$ can be expressed as
\vspace{-0.1cm}
\begin{equation}
\overline{\textbf{h}}_{ai}=[1,e^{j2\pi\frac{d}{\lambda}\text{sin}\varsigma_i},\ldots,e^{j2\pi\frac{d}{\lambda}(L-1)\text{sin}\varsigma_i}]^T,
\end{equation}
\begin{equation}
\vspace{-0.2cm}
\overline{\textbf{h}}_{bi}=[1,e^{j2\pi\frac{d}{\lambda}\text{sin}\varphi_i},\ldots,e^{j2\pi\frac{d}{\lambda}(L-1)\text{sin}\varphi_i}]^T,
\end{equation}
where $\varphi_i$ and $\varsigma_i$ represent the $i$th pair of users' angle of arrival (AoA) and angle of departure (AoD), respectively. In the present paper, we consider RISs that are made of discrete elements that are spaced half of the wavelength apart \cite{9140329}. Therefore, we assume $d=\frac{\lambda}{2}$ in the rest of the paper.

%Let $p_i$ denotes the power transmitted by $\text U_{Ai}$.
%$$\textbf{h}_{ai} %\sim\mathcal{CN}(\textbf{0},\sigma_{gi}^2\textbf{I}_L)$$

%$x_{Ai}[t]$ has not be decided yet.
%The unit-energy information symbols from U1 and U2, randomly selected from the codebook, are denoted by s1 and s2, respectively.

%\subsection{perfect CSI}
We assume the availability of statistical CSI at $\text U_{Ai}$, for $i = 1, \ldots,K$.
The statistical CSI can be readily obtained since it varies much slowly than the instantaneous CSI.
%The channel sate information can be acquired similar to \cite{taha2019enabling}.
The signal received at $\text U_{Bi}$ is given by
\begin{align}\label{y_i1}
y_{i}&= \textbf{g}_{bi}^T\bm{\Theta}{\sum_{j=1}^K \sqrt{p_{j}} \textbf{g}_{aj} x_{j}}+n_{i}
\nonumber\\
&= \underbrace{\sqrt{p_{i}} \textbf{g}_{bi}^T\bm{\Theta} \textbf{g}_{ai} x_{i}}_{\text{Desired\ signal}}+\underbrace {\sum_{j=1,j\not=i}^K \sqrt{p_{j}} \textbf{g}_{bi}^T\bm{\Theta} \textbf{g}_{aj} x_{j}}_{\text{Inter-user\ interference}} + \underbrace {n_{i}}_{\text{Noise}},
\end{align}
%\end{figure*}
where $x_{j}$ $\sim \mathcal{CN}(0,1)$ represents the signal transmitted by $\text U_{Aj}$,
$p_{j}$ denotes the transmit power of $\text U_{Aj}$, and
$n_{i}$ $\sim \mathcal{CN}(0,\sigma_{i}^2)$ is the additive white Gaussian noise (AWGN) at $\text U_{Bi}$, for $i = 1,\ldots,K$.

From \eqref{y_i1}, we observe that $y_{i}$ consists of three parts: the desired signal that $\text U_{Bi}$ wants to receive, the interference produced by other multi-pair users, and
the noise.
Furthermore, the signal-to-interference
plus noise ratio (SINR) at $\text U_{Bi}$ is given by
\vspace{-0.2cm}
\begin{equation}\label{SINR_i}
\vspace{-0.1cm}
{\rm{\gamma}}_{i}= \frac{p_{i}\alpha_{bi} \alpha_{ai}\left|\textbf{h}_{bi}^T\bm{\Theta}\textbf{h}_{ai}\right|^2}{\sum\limits_{j=1,j\not =i}^K\left(p_{j}\alpha_{bi} \alpha_{aj}\left|\textbf{h}_{bi}^T\bm{\Theta}\textbf{h}_{aj}\right|^2\right)+\sigma_{i}^2}.
\end{equation}

Hence, the ergodic achievable rate for $\text U_{Bi}$ can be expressed as
\vspace{-0.2cm}
\begin{equation}\label{R_i1}
\vspace{-0.2cm}
R_{i}=\mathbb{E}\{\text{log}_2(1+{\rm{\gamma}}_{i})\},
\end{equation}

and the sum achievable rate can be written as
\vspace{-0.1cm}
\begin{equation}\label{C}
\vspace{-0.3cm}
C=\sum_{i=1}^K R_{i}.
\end{equation}

%\section{Performance Optimization Analysis}

%We derive closed-form approximate expression for the achievable rate.
\section{Achievable Rate Analysis}
In this section, we analyze the system performance.
%To this end, we introduce the following theorem.
We derive an approximate expression for the achievable rate in the following theorem.
%To analyze the performance of RIS-aided multi-pair communication systems, we firstly

\begin{theorem}\label{thm_limited}
In an RIS-aided multi-pair communication system, the average achievable rate of $\text U_{Bi}$ can be approximated as
\begin{align}\label{Ri3}
R_i \! \approx  \!  \tilde{R}_i \triangleq \text{log}_2  \! \left( \! 1+\frac{p_{i}\alpha_{bi} \alpha_{ai}\frac{\varepsilon_i\beta_i\Omega_{i,i}+L(\varepsilon_i+\beta_i)+L}{(\varepsilon_i+1)(\beta_i+1)}}{\sum\limits_{\mathclap{j=1,j\not =i}}^K  \left(p_{j}\alpha_{bi} \alpha_{aj}\frac{\varepsilon_i\beta_j\Omega_{i,j}+L(\varepsilon_i+\beta_j)+L}{(\varepsilon_i+1)(\beta_j+1)}\right)+\sigma_{i}^2}\right) ,
\end{align}
where $\Omega_{i,i}$ and $\Omega_{i,j}$ are, respectively, defined as
\vspace{-0.1cm}
\begin{equation}\label{oij0}
\vspace{-0.0cm}
\!\!\!\!\Omega_{i,i}
\!=\!L+2\sum_{\mathclap{1\leq m<n\leq L}}
\text{cos}[\theta_n\!-\!\theta_m+(n-m)\pi(\text{sin}\varphi_i\!+\!\text{sin}\varsigma_i)],
\end{equation}
\begin{equation}\label{oij1}
\vspace{-0.1cm}
\!\!\!\!\Omega_{i,j}
\!=\!L+2\sum_{\mathclap{1\leq m<n\leq L}}
\text{cos}[\theta_n\!-\!\theta_m+(n-m)\pi(\text{sin}\varphi_i\!+\!\text{sin}\varsigma_j)].\!\!
\end{equation}
%\hrulefill
%\end{figure*}
\end{theorem}

\begin{figure*}[b]
\vspace{-0.4cm}
\hrulefill
\setcounter{equation}{17}
\begin{align}\label{htgr}
\mathfrak{R}\left(\textbf{h}_{bi}^T\bm{\Theta}\textbf{h}_{aj}\right)
&
=\lambda_{ij} \sum_{\ell=1}^L  \left( \right.\sqrt{\varepsilon_i\beta_j}
\left(
\text{cos}(\theta_\ell +(\ell-1)\pi(\text{sin}\varphi_i+\text{sin}\varsigma_j))
\right)+\text{cos}\theta_\ell (s_{\ell i}u_{\ell j}-t_{\ell i}v_{\ell j})
-\text{sin}\theta_\ell (s_{\ell i}v_{\ell j}+t_{\ell i}u_{\ell j})\left.\right)
\nonumber\\
&
+\lambda_{ij} \sum_{\ell=1}^L \left( \right.
\sqrt{\varepsilon_i}
\left(
(\text{cos}\theta_\ell -\text{sin}\theta_\ell)\text{cos}((\ell-1)\pi \text{sin}\varphi_i) u_{\ell j}
-(\text{cos}\theta_\ell + \text{sin}\theta_\ell)\text{sin}((\ell-1)\pi \text{sin}\varphi_i) v_{\ell j}
\right)
\nonumber\\
&
+\sqrt{\beta_j}
\left(
(\text{cos}\theta_\ell -\text{sin}\theta_\ell) (\text{cos}((\ell-1)\pi \text{sin}\varsigma_j) s_{\ell i}
-(\text{cos}\theta_\ell + \text{sin}\theta_\ell)\text{sin}((\ell-1)\pi \text{sin}\varsigma_j) t_{\ell i})
\right)
\left.\right)
%\vspace{-0.8cm}
\end{align}
%\normalsize
\vspace{-0.1cm}
\hrulefill
\vspace{-0.1cm}
\setcounter{equation}{18}
\begin{align}\label{htgi}
\mathfrak{I}\left(\textbf{h}_{bi}^T\bm{\Theta}\textbf{h}_{aj}\right)
&
= \lambda_{ij} \sum_{\ell=1}^L
\left( \right.
\sqrt{\varepsilon_i\beta_j}
\left(
\text{sin}(\theta_\ell+(\ell-1)\pi(\text{sin}\varphi_i+\text{sin}\varsigma_j))
\right)+\text{sin}\theta_\ell (s_{\ell i}u_{\ell j}-t_{\ell i}v_{\ell j})
+\text{cos}\theta_\ell (s_{\ell i}v_{\ell j}+t_{\ell i}u_{\ell j})
\left.\right)
\nonumber\\
&
+\lambda_{ij} \sum_{\ell=1}^L
\left( \right.
\sqrt{\varepsilon_i}
\left(
(\text{cos}\theta_\ell + \text{sin}\theta_\ell) \text{cos}((\ell-1)\pi \text{sin}\varphi_i) u_{\ell j}
+(\text{cos}\theta_\ell - \text{sin}\theta_\ell)\text{sin}((\ell-1)\pi \text{sin}\varphi_i) v_{\ell j}
\right)
\nonumber\\
&
+\sqrt{\beta_i}
\left(
(\text{cos}\theta_\ell + \text{sin}\theta_\ell) \text{cos}((\ell-1)\pi \text{sin}\varsigma_j) s_{\ell j}
+(\text{cos}\theta_\ell - \text{sin}\theta_\ell)\text{sin}((\ell-1)\pi \text{sin}\varsigma_j) t_{\ell j}
\right)
\left.\right)
\vspace{-0.05cm}
\end{align}
\normalsize
\setcounter{equation}{13}
%\hrulefill
\vspace{-0.6cm}
\end{figure*}

\itshape \textbf{Proof:}
\upshape
%We aim to prove \emph{Proposition}~\ref{thm_limited}.
Using Lemma~1 in \cite{zhang2014power}, $R_i$ in \eqref{R_i1} can be approximated as
\vspace{-0.3cm}
\begin{equation}\label{Ri2}
%\vspace{-0.3cm}
R_i \! \approx  \!
\text{log}_2\left(1+\frac{p_{i}\alpha_{bi} \alpha_{ai}\mathbb{E}\left\{\left|\textbf{h}_{bi}^T\bm{\Theta}\textbf{h}_{ai}\right|^2\right\}}{\sum\limits_{j=1,j\not =i}^K\left(p_{j}\alpha_{bi} \alpha_{bj}\mathbb{E}\left\{\left|\textbf{h}_{bi}^T\bm{\Theta}\textbf{h}_{aj}\right|^2\right\}\right)+\sigma_{i}^2}\right).
\end{equation}

The $\ell$th element of $\textbf{h}_{ai}$ and the $\ell$th element of $\textbf{h}_{bi}$ can be  written as
\vspace{-0.2cm}
\begin{equation}\label{hli}
\vspace{-0.1cm}
[\textbf{h}_{ai}]_\ell=\sqrt{\frac{\varepsilon_i}{\varepsilon_i+1}}e^{j(\ell-1)\pi \text{sin}\varphi_i}+\sqrt{\frac{1}{\varepsilon_i+1}} (s_{\ell i}+j t_{\ell i}),
\end{equation}
\begin{equation} \label{glj}
\vspace{-0.2cm}
[\textbf{h}_{bi}]_\ell=\sqrt{\frac{\beta_i}{\beta_i+1}}e^{j(\ell-1)\pi \text{sin}\varsigma_i}+\sqrt{\frac{1}{\beta_i+1}} (u_{\ell i}+j v_{\ell i}),
\end{equation}
where $s_{\ell i} \sim \mathcal{N}(0,1/2)$ and $t_{\ell i} \sim \mathcal{N}(0,1/2)$ denote the independent real and imaginary parts of $[\tilde{\textbf{h}}_{ai}]_\ell$, respectively, and
$u_{\ell j} \sim \mathcal{N}(0,1/2)$ and $v_{\ell j} \sim \mathcal{N}(0,1/2)$ denote the independent real and imaginary parts of $[\tilde{\textbf{h}}_{bi}]_\ell$, respectively.

Let us first consider the term $\mathbb{E}\left\{\left|\textbf{h}_{bi}^T\bm{\Theta}\textbf{h}_{aj}\right|^2\right\}$ in \eqref{Ri2}.
By substituting \eqref{hli} and \eqref{glj} into $\mathbb{E}\left\{\left|\textbf{h}_{bi}^T\bm{\Theta}\textbf{h}_{aj}\right|^2\right\}$, it can be rewritten as
\vspace{-0.1cm}
\begin{equation}\label{htg2}
\vspace{-0.1cm}
\!\!\mathbb{E}\left\{\left|\textbf{h}_{bi}^T\bm{\Theta}\textbf{h}_{aj}\right|^2\right\}\!=\!
\mathbb{E}\left\{\mathfrak{R}\left(\textbf{h}_{bi}^T\bm{\Theta}\textbf{h}_{aj}\right)^2
+\mathfrak{I}\left(\textbf{h}_{bi}^T\bm{\Theta}\textbf{g}_{aj}\right)^2\right\},\!\!\!
\end{equation}
where the terms $\mathfrak{R}\left(\textbf{h}_{bi}^T\bm{\Theta}\textbf{h}_{aj}\right)$ and $\mathfrak{I}\left(\textbf{h}_{bi}^T\bm{\Theta}\textbf{g}_{aj}\right)$ are the real and imaginary parts of $\textbf{h}_{bi}^T\bm{\Theta}\textbf{h}_{aj}$.
They are respectively given by \eqref{htgr} and \eqref{htgi} (at the bottom of this page), where $\lambda_{ij}$ is given by
\vspace{-0.1cm}
\setcounter{equation}{19}
\begin{equation}
\vspace{-0.1cm}
\lambda_{ij} = \frac{1}{\sqrt{(\varepsilon_i+1)(\beta_j+1)}}.
\end{equation}
With the aid of some algebraic calculations, we obtain
\vspace{-0.1cm}
\begin{equation}\label{E-htheta-ij}
\vspace{-0.1cm}
\mathbb{E}\left\{\left|\textbf{h}_{bi}^T\bm{\Theta}\textbf{h}_{aj}\right|^2\right\}=
\frac{\varepsilon_i\beta_j\Omega_{i,j}+L(\varepsilon_i+\beta_j)+L}{(\varepsilon_i+1)(\beta_j+1)}.
\end{equation}

A similar method can be used to calculate the term $\mathbb{E}\left\{\left|\textbf{h}_{bi}^T\bm{\Theta}\textbf{h}_{ai}\right|^2\right\}$ in \eqref{Ri2}, which yields
\vspace{-0.1cm}
\begin{equation}\label{E-htheta-ii}
\vspace{-0.1cm}
\mathbb{E}\left\{\left|\textbf{h}_{bi}^T\bm{\Theta}\textbf{h}_{ai}\right|^2\right\}=
\frac{\varepsilon_i\beta_i\Omega_{i,i}+L(\varepsilon_i+\beta_i)+L}{(\varepsilon_i+1)(\beta_i+1)}.
\end{equation}

By substituting \eqref{E-htheta-ij} and \eqref{E-htheta-ii} into \eqref{Ri2}, we obtain the final result in \eqref{Ri3}. This completes the proof.
\hfill $\Box$

Substituting \eqref{Ri3} into \eqref{C}, we obtain the sum achievable rate. According to Theorem~\ref{thm_limited}, if $\alpha_{bi}$, $\alpha_{ai}$, AoA, AoD, $\varepsilon_i$, and $\sigma_i$  are kept fixed, the sum achievable rate is determined by the number of user pairs $K$, the transmit power $p_i$, the phase shift matrix $\bm{\Theta}$ and
the number of reflective elements $L$.

\section{Phase Shifts Optimization}
In this section, we formulate and solve the phase shifts optimization problem in order to maximize the sum achievable rate. Both case studies of CPSs and DPSs are considered.

\subsection{Optimal CPSs Design}
To begin with, we consider the scenario adopting CPSs.
The optimization problem is formulated as
\begin{subequations}\label{p}
\begin{alignat}{1}
\max_{\bm{\Theta}} \quad
&{\sum_{i=1}^K  \tilde{R}_i}  \\
\mbox{s.t.}\quad
&\theta_\ell\in [0,2 \pi)\ \forall \ell = 1,\ldots,L.
\end{alignat}
\end{subequations}

\renewcommand{\algorithmcfname}{Algorithm}
%\vspace{-0.6cm}
\begin{algorithm}
 \caption{Genetic Algorithm}\label{ga}
Initialize the iteration number $n$ = 1, $f_{min}^1$ = 1, and generate the initial population\;
\While{$n \leq n_{max}$ or $f_{min}^n > \varepsilon$}{
Apply evaluation and sort\;
\While{The number of children does not reach $N_{\rm{c}}$}{
Apply the selection function\;
Apply the crossover function\;
}
Place the first $N_{\rm{s}}$ individuals in the list, and the $N_{\rm{c}}$ children in the current generation\;
Select $N_{\rm{e}}$ elites who have the lowest fitness function value\;
Apply the mutation function\;
Refresh the current generation, with elites and mutated
individuals\;
Return the lowest fitness function value in the current generation $f_{min}^n$\;
Set $n\leftarrow n+1$\;
}
Output the individual with the lowest fitness function value in the current generation.
\end{algorithm}
%\vspace{-0.2cm}

Problem \eqref{p} can hardly be solved by the conventional methods, such as successive convex approximation, minorization-maximization, semidefinite program, second-order cone program, etc.

To tackle this issue, we propose a GA method in Algorithm \ref{ga}.
We associate each individual to a  1 $\times$ $L$ phase shift vector $\bm\theta$ = $[\theta_1,\ldots,\theta_\ell,\ldots,\theta_L ]$ and $\theta_\ell$ corresponds to its $\ell$th gene.

The initial population, the evaluation and sort operations, the selection function, the crossover function and the mutation function of the proposed GA method are described as follows.

1) Initial population:
$N_{\rm{t}}$ individuals are generated, by generating each gene as a variable that is randomly distributed within $[0, 2\pi)$. This is referred to as the initial population.

2) Evaluation and sort:
Each individual is evaluated by using the fitness function
\vspace{-0.1cm}
\begin{equation}\label{fthat}
\vspace{-0.0cm}
f(\bm\theta)=\frac{1}{\sum_{i=1}^K  \tilde{R}_i}.
\end{equation}
In particular, the lower the fitness function is, the higher the evaluation of the individual and the position in the corresponding sorted list are.

3) Selection function:
The selection function is used to
obtain two candidates, $\bm\theta_{\rm{1}}$ and $\bm\theta_{\rm{2}}$, from $N_{\rm{t}}$ individuals.
The individuals with larger values of the fitness function have a lower probability to be selected.
$\bm\theta_{\rm{1}}$ and $\bm\theta_{\rm{2}}$ are obtained by using Algorithm \ref{selection}.

\renewcommand{\algorithmcfname}{Algorithm}
\begin{algorithm}
 \caption{Selection Algorithm}\label{selection}
  Generate a random number $c$ between 0 and 1\;
  Define a cumulative-probabilities vector $\bm{r} = [1/N_{\rm{t}},\ldots,r/N_{\rm{t}},\ldots,1]$, for $r = 1, . . . , N_{\rm{t}}$.\;
  Denote $r^\prime/N_{\rm{t}}$ as the nearest maximum number in $\bm{r}$ to $c$\;
  Take the $r^\prime$th individual in the list that is sorted by using \eqref{fthat}.
\end{algorithm}

\renewcommand{\algorithmcfname}{Algorithm}
\begin{algorithm}%[t]
 \caption{Single-Point Crossing Algorithm}\label{crossover}
%  \While{The number of children does not reach $N_{\rm{c}}$}{
  Get $\bm\theta_{\rm{1}}=[\theta_1^{(1)},\ldots,\theta_\ell^{(1)},\ldots,\theta_L^{(1)} ]$ and $\bm\theta_{\rm{2}}=[\theta_1^{(2)},\ldots,\theta_\ell^{(2)},\ldots,\theta_L^{(2)} ]$ selected by Algorithm \ref{selection}\;
  Identify a crossover point $\ell^\prime \in [1,L]$ randomly\;
  Crossover $\bm\theta_{\rm{1}}$ and $\bm\theta_{\rm{2}}$ at crossover point $\ell^\prime$\;
  Obtain two children $\bm\theta_{\rm{c1}}=[\theta_1^{(1)},\ldots,\theta_{\ell^\prime}^{(1)},
  \theta_{\ell^\prime+1}^{(2)},\ldots,\theta_L^{(2)} ]$ and $\bm\theta_{\rm{c2}}=[\theta_1^{(2)},\ldots,\theta_{\ell^\prime}^{(2)} ,\theta_{\ell^\prime+1}^{(1)},\ldots,\theta_L^{(1)}]$.
 %}
\end{algorithm}

4) Crossover function:
The crossover function operates on $\bm\theta_{\rm{1}}$ and $\bm\theta_{\rm{2}}$, and generates two crossover children $\bm\theta_{\rm{c1}}$ and $\bm\theta_{\rm{c2}}$.
A single-point crossing algorithm is employed, which is described in Algorithm \ref{crossover}.

5) Mutation function:
The mutation function operates on $(N_{\rm{t}}$-$N_{\rm{e}})L$ genes from $(N_{\rm{t}}$-$N_{\rm{e}})$ individuals (except for the elites).
Each gene is capable of mutating to a random number between 0 and $2\pi$ with mutation rate.

Similar as \cite{8269405}, the complexity of the proposed GA algorithm is $n N_t$, where $N_t$ is the population size, and $n$ is the number of generations evaluated. Moreover, $n$ is determined by the convergence behavior of the GA.

\subsection{Optimal DPSs Design}

In practice, only a limited number of phase shifts can be used.
We assume that each reflective element is encoded with $B$ bits, and thus $2^B$ phase shifts can be chosen to enhance the signal reflected by the RIS.
We denote the DPS matrix as $\bm{\hat{\Theta}}=diag(e^{j\hat{\theta}_1},\ldots,e^{j\hat{\theta}_\ell},\ldots,e^{j\hat{\theta}_L})$, where $\hat{\theta}_\ell$ is the DPS of the $\ell$th reflective element.
Replacing the CPSs in Problem \eqref{p} by the DPSs, the optimization problem for DPSs can be formulated as
\begin{subequations}\label{p2}
\begin{alignat}{1}
\max_{\bm{\hat{\Theta}}} \quad
& {\sum_{i=1}^K  {\tilde{R}_i }}   \\
\mbox{s.t.}\quad
&\hat{\theta}_\ell\in \{ 0,2\pi/2^B,\ldots,2\pi(2^B-1)/2^B \}
\nonumber\\
&\ \forall \ell = 1, \ldots,L.
\end{alignat}
\end{subequations}

It is observed that Problem \eqref{p2} is similar to Problem \eqref{p}. Accordingly, the GA method proposed for solving the CPSs optimization problem can be applied to solve the DPSs optimization problem as well.

%begin table
\renewcommand\arraystretch{1.5}  %行距
\renewcommand\tabcolsep{15.0 pt} % 调整表格列间的长度
\begin{table}[t]
\vspace{-0.4cm}
\caption{Parameters for Simulation}
\vspace{-0.4cm}
\label{tab:1}       % Give a unique label
%\vspace{-0.4cm}
\begin{center}
%\resizebox{\textwidth}{12mm}
%\resizebox{\textwidth}{18mm}
\begin{tabular}{|c|c|c|c|}
%\begin{tabular}{lll}
 \hline
{\bfseries No. of pairs}  & \bfseries {AoA} & \bfseries {AoD} & \bfseries $\alpha_{ai}$ and $\alpha_{bi}$\\
 \hline
 {\bfseries 1}  & 5.5629 & 1.1450 & 0.0023\\
 {\bfseries 2}  & 5.6486 & 0.6226 & 0.0285 \\
 {\bfseries 3}  & 3.9329 & 3.0773 & 0.0025 \\
 {\bfseries 4}  & 0.8663 & 1.2142 & 0.0012 \\
 {\bfseries 5}  & 1.3685 & 5.6290 & 0.0550 \\
 {\bfseries 6}  & 1.1444 & 0.6226 & 0.0141 \\
\hline
\end{tabular}
\end{center}
\vspace{-0.4cm}
\end{table}
%end table

\section{ Numerical Results}

We evaluate the impact of different parameters on the sum achievable rate. We assume that the Rician factor is $\varepsilon_i$ = 10, the noise power is $\sigma_i^2 = 1$, and the transmission power is SNR = $p_i$, for $i = 1, . . . , K$. The main parameters for the GA are: $N_{\rm{t}}$ = 100, $N_{\rm{s}}$ = 50, $N_{\rm{c}}$ = 50, $N_{\rm{e}}$ = 1, $n_{max}$ = 10000, $P_{\rm{m}}$ = 0.1, and $\varepsilon$ = $10^{-6}$.
The other parameters are summarized in Table \ref{tab:1}, where the AOA and the AOD are randomly distributed within [0, 2$\pi$), and the large-scale fading coefficients $\alpha_{ai}$ and $\alpha_{bi}$ are set according to \cite{zhang2014power}.

\begin{figure}[t]
%\vspace{-0.6cm}
\centering
\includegraphics[scale=0.65]{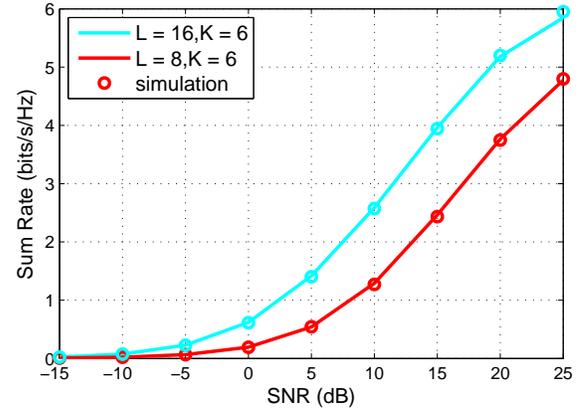}
\vspace{-0.2cm}
\caption{Sum achievable rate versus SNR with $B = 2$ and GA. } \label{fig1}
\vspace{-0.4cm}
\end{figure}

In Fig. \ref{fig1}, we illustrate the sum achievable rate \eqref{C} versus the SNR obtained from the analytical expression in \eqref{Ri3} and Monte-Carlo simulations from  \eqref{R_i1}. Two quantization bits are assumed. A good analytical agreement with Monte-Carlo simulation results is obtained, which verifies the derivations.
We observe that the sum achievable rate increases with the number of reflecting elements $L$, as expected.

\begin{figure}[t]
\vspace{-0.55cm}
\centering
\includegraphics[scale=0.65]{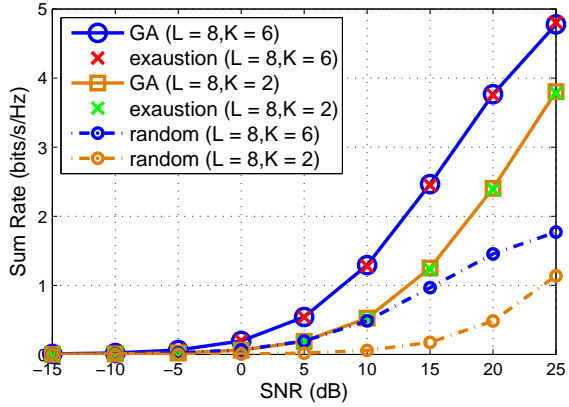}
\vspace{-0.15cm}
\caption{Sum achievable rate versus SNR with $B = 2$ by various schemes.} \label{fig2}
\vspace{-0.4cm}
\end{figure}

Fig. \ref{fig2} depicts the sum achievable rate versus the SNR, by assuming two quantization bits and comparing different optimization schemes. Compared with the scheme based on randomly chosen phase shifts, the proposed GA and exhaustive search methods can achieve higher sum achievable rate. It is interesting to observe that the proposed GA method has almost the same performance as the globally optimal solution obtained by using the exhaustive search method.

\begin{figure}[t]
%\vspace{-0.6cm}
\centering
\includegraphics[scale=0.65]{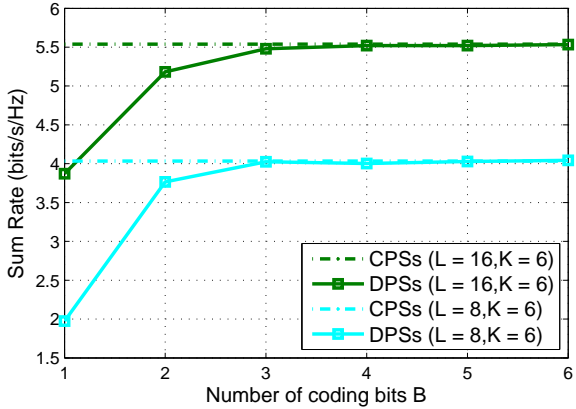}
\vspace{-0.15cm}
\caption{Sum achievable rate versus $B$ with SNR $= 20$ dB. } \label{fig3}
\vspace{-0.4cm}
\end{figure}
%Let us discuss the influence of the limited phase shifts on the data rate.

Fig. \ref{fig3} shows the sum achievable rate versus $B$ (the number of quantization) for the scenarios adopting DPSs.
The sum achievable rate for the DPSs setup increases rapidly when $B$ is small, while the curve gradually saturates when $B$ is large. However, it is  known that using a large number of quantization bits to control the phase shifts incurs high hardware cost and power consumption.
The figure shows that three quantization bits can achieve a large portion of the sum achievable rate for the CPSs setup,
which provides useful engineering design insights for RIS-aided systems.

\begin{figure}[t]
\vspace{-0.5cm}
\centering
\includegraphics[scale=0.65]{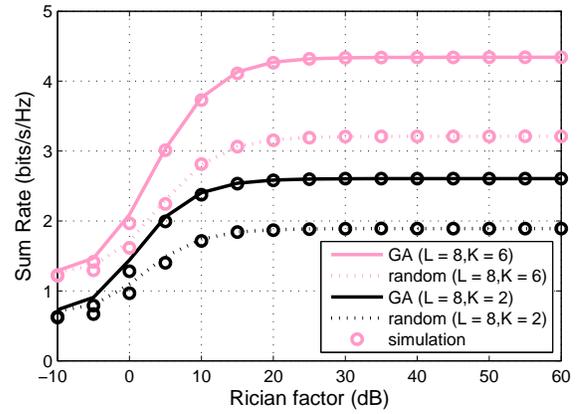}
\vspace{-0.2cm}
      \caption{Sum achievable rate versus Rician factor with $B$ = 2 and SNR = 20 dB.} \label{figK}
\vspace{-0.4cm}
\end{figure}

Fig. \ref{figK} depicts the sum achievable rate versus the Rician factor, with two quantization bits and SNR = 20 dB. The figure shows that the sum achievable rate increases with the Rician factor.
With the increase of the Rician factor, we can observe that the gaps of the sum rates between the random scheme and the proposed GA method converge to the fixed values.
It is because with a large Rician factor, the channels are dominated by the LOS component.
%It shows that phase shifts have a little influence on the NLoS component of the channel, while the LoS component would be changed significantly by the phase shift $\Theta$.

%\vspace{-0.2cm}
\section{Conclusion}
In this paper, we investigated RIS-aided communications for multiple pairs of users.
We derived an approximate expression for the achievable rate.
Based on the derived analytical framework, we developed a GA-based algorithm for optimizing the achievable sum achievable rate, which can be applied to CPS-based and DPS-based implementations.
%Simulation results verified the correctness of the proposed GA-based method.
Simulation results verified the effectiveness of the proposed GA-based method.

%\appendices
%\section{}\label{applimited}

%
%%
%%
%%
%%%\section*{Acknowledgment}
%%
%%
%%%The authors would like to thank...
%%
%%
%%
%%\ifCLASSOPTIONcaptionsoff
%%  \newpage
%%\fi

%%\begin{thebibliography}{1}
%%% You can use other form of bib file by changing here...
%%
%%\bibitem{Cui2014Coding}
%%E.~Basar, M.~Di~Renzo, J.~de~Rosny, M.~Debbah, M.-S. Alouini, and R.~Zhang,
%%  ``Coding metamaterials, digital metamaterials and programmable metamaterials,''
%%  \emph{Light-Science \& Applications}, 2019.
%%
%%\end{thebibliography}

%\vspace{-0.2cm}

\bibliographystyle{IEEEtran}
\bibliography{IEEEabrv,Ref}

\begin{thebibliography}{10}
\providecommand{\url}[1]{#1}
\csname url@samestyle\endcsname
\providecommand{\newblock}{\relax}
\providecommand{\bibinfo}[2]{#2}
\providecommand{\BIBentrySTDinterwordspacing}{\spaceskip=0pt\relax}
\providecommand{\BIBentryALTinterwordstretchfactor}{4}
\providecommand{\BIBentryALTinterwordspacing}{\spaceskip=\fontdimen2\font plus
\BIBentryALTinterwordstretchfactor\fontdimen3\font minus
  \fontdimen4\font\relax}
\providecommand{\BIBforeignlanguage}[2]{{%
\expandafter\ifx\csname l@#1\endcsname\relax
\typeout{** WARNING: IEEEtran.bst: No hyphenation pattern has been}%
\typeout{** loaded for the language `#1'. Using the pattern for}%
\typeout{** the default language instead.}%
\else
\language=\csname l@#1\endcsname
\fi
#2}}
\providecommand{\BIBdecl}{\relax}
\BIBdecl

\bibitem{9140329}
M.~D. {Renzo} \emph{et~al.}, ``Smart radio environments empowered by
  reconfigurable intelligent surfaces: How it works, state of research, and
  road ahead,'' \emph{IEEE J. Sel. Areas Commun.}, early access, July 2020,
  doi:{\color{blue} 10.1109/JSAC.2020.3007211}.

\bibitem{8741198}
C.~{Huang}, A.~{Zappone}, G.~C. {Alexandropoulos}, M.~{Debbah}, and C.~{Yuen},
  ``Reconfigurable intelligent surfaces for energy efficiency in wireless
  communication,'' \emph{IEEE Trans. Wireless Commun.}, vol.~18, no.~8, pp.
  4157--4170, Aug. 2019.

\bibitem{8811733}
Q.~{Wu} and R.~{Zhang}, ``Intelligent reflecting surface enhanced wireless
  network via joint active and passive beamforming,'' \emph{IEEE Trans.
  Wireless Commun.}, vol.~18, no.~11, pp. 5394--5409, Nov. 2019.

\bibitem{2019arXiv191112296H}
\BIBentryALTinterwordspacing
C.~{Huang} \emph{et~al.}, ``{Holographic MIMO surfaces for 6G wireless
  networks: Opportunities, Challenges, and Trends}.'' [Online]. Available:
  \url{https://arxiv.org/abs/1911.12296v1}
\BIBentrySTDinterwordspacing

\bibitem{2019arXiv190308925D}
M.~{Di Renzo} \emph{et~al.}, ``{Smart radio environments empowered by
  recongurable AI meta-surfaces: An idea whose time has come},'' \emph{EURASIP
  J. Wireless Commun. Netw.}, vol. 2019, p. 129, May 2019.

\bibitem{Shen2019Secrecy}
H.~Shen, W.~Xu, S.~Gong, Z.~He, and C.~Zhao, ``Secrecy rate maximization for
  intelligent reflecting surface assisted multi-antenna communications,''
  \emph{IEEE Commun. Lett.}, vol.~23, no.~9, pp. 1488--1492, Sep. 2019.

\bibitem{2019arXiv191201497Y}
\BIBentryALTinterwordspacing
X.~{Yu}, D.~{Xu}, Y.~{Sun}, D.~W.~K. {Ng}, and R.~{Schober}, ``{Robust and
  secure wireless communications via intelligent reflecting surfaces}.''
  [Online]. Available: \url{https://arxiv.org/abs/1912.01497}
\BIBentrySTDinterwordspacing

\bibitem{9090356}
C.~{Pan}, H.~{Ren}, K.~{Wang}, W.~{Xu}, M.~{Elkashlan}, A.~{Nallanathan}, and
  L.~{Hanzo}, ``Multicell {MIMO} communications relying on intelligent
  reflecting surfaces,'' \emph{IEEE Trans. Wireless Commun.}, early access, May
  2020, doi:{\color{blue} 10.1109/TWC.2020.2990766}.

\bibitem{200605147}
\BIBentryALTinterwordspacing
Z.~{Peng}, Z.~{Zhang}, C.~{Pan}, L.~{Li}, and A.~L. {Swindlehurst},
  ``{Multiuser full-duplex two-way communications via intelligent reflecting
  surface}.'' [Online]. Available: \url{https://arxiv.org/abs/2006.05147}
\BIBentrySTDinterwordspacing

\bibitem{Tong2020Latency}
\BIBentryALTinterwordspacing
T.~Bai, C.~Pan, Y.~Deng, M.~Elkashlan, and L.~Hanzo, ``Latency minimization for
  intelligent reflecting surface aided mobile edge computing.'' [Online].
  Available: \url{https://arxiv.org/abs/1910.07990}
\BIBentrySTDinterwordspacing

\bibitem{2019arXiv190804863P}
C.~{Pan} \emph{et~al.}, ``Intelligent reflecting surface aided {MIMO}
  broadcasting for simultaneous wireless information and power transfer,''
  \emph{IEEE J. Sel. Areas Commun.}, early access, June 2020, doi:{\color{blue}
  10.1109/JSAC.2020.3000802}.

\bibitem{9106840}
J.~{Xu}, Y.~{Wang}, W.~{Xu}, S.~{Jin}, H.~{Shen}, and X.~{You}, ``On uplink
  performance of multiuser massive {MIMO} relay network with limited {RF}
  chains,'' \emph{IEEE Trans. Veh. Technol.}, early access, June 2020,
  doi:{\color{blue} 10.1109/TVT.2020.2999345}.

\bibitem{7947159}
W.~{Xu}, J.~{Liu}, S.~{Jin}, and X.~{Dong}, ``Spectral and energy efficiency of
  multi-pair massive {MIMO} relay network with hybrid processing,'' \emph{IEEE
  Trans. Commun.}, vol.~65, no.~9, pp. 3794--3809, Sep. 2017.

\bibitem{2019arXiv190808747D}
\BIBentryALTinterwordspacing
M.~{Di Renzo} \emph{et~al.}, ``{Reconfigurable intelligent surfaces vs.
  relaying: Differences, similarities, and performance comparison}.'' [Online].
  Available: \url{https://arxiv.org/abs/1908.08747}
\BIBentrySTDinterwordspacing

\bibitem{9005237}
C.~{Guo}, Y.~{Cui}, F.~{Yang}, and L.~{Ding}, ``Outage probability analysis and
  minimization in intelligent reflecting surface-assisted {MISO} systems,''
  \emph{IEEE Commun. Lett.}, early access, Feb. 2020, doi:{\color{blue}
  10.1109/LCOMM.2020.2975182}.

\bibitem{zhang2014power}
Q.~Zhang, S.~Jin, K.-K. Wong, H.~Zhu, and M.~Matthaiou, ``Power scaling of
  uplink massive {MIMO} systems with arbitrary-rank channel means,'' \emph{IEEE
  J. Sel. Topics Signal Process.}, vol.~8, no.~5, pp. 966--981, Oct. 2014.

\bibitem{8269405}
Z.~{Ye}, C.~{Pan}, H.~{Zhu}, and J.~{Wang}, ``Tradeoff caching strategy of the
  outage probability and fronthaul usage in a cloud-{RAN},'' \emph{IEEE Trans.
  Veh. Technol.}, vol.~67, no.~7, pp. 6383--6397, Jan. 2018.

\end{thebibliography}

\iffalse
\begin{IEEEbiography}{Yuguang ``Michael'' Fang}
Biography text here.
\end{IEEEbiography}
\fi

%It is not necessary to upload the biography when you submit your manuscript.

\end{document}